# The dissolution of a miscible drop rising or falling in another liquid at low Reynolds number

Jan Martin Nordbotten[1]
Center for Modeling of Coupled Subsurface Dynamics
Department of Mathematics
University of Bergen
Norway

Endre Joachim Lerheim Mossige
RITMO Centre for Interdisciplinary Studies in Rhythm, Time and Motion
University of Oslo
Norway

1: Corresponding author (Jan.Nordbotten@uib.no)

## Abstract

"A basic and basically unsolved problem in fluid dynamics is to determine the evolution of rising bubbles and falling drops of one miscible liquid in another"[1]. Here, we address this important literature gap and present the first theory predicting the velocity, volume and composition of such drops at low Reynolds numbers. For the case where the diffusion out of the drop is negligible, we obtain a universal scaling law. For the more general case where diffusion occurs into and out of the drop, the full dynamics is governed by a parameter-free first-order ordinary differential equation, whose closed form solution exists, and only depends on the initial condition. Our analysis depends primarily on "drop-scale" effective parameters for the diffusivity through the interfacial boundary layer. We validate our results against experimental data for water drops suspended in syrup, corresponding to certain regimes of the mass exchange ratio between water and syrup, and by this explicitly identify the drop-scale parameters of the theory.

# 1. Introduction

The study of drops and bubbles is at the heart of numerous problems in fluid mechanics[2,3,4] and can be approached with simple and affordable ingredients[5,6]. For example, we can learn about capillarity and surface tension simply by watching a coffee drop dry[7,8], we can visualize the beautiful dynamics of rising bubbles[9] when we drink a carbonated beverage, and when we sweeten our tea, we can investigate the fascinating dynamics of mixing[10].

When a viscous drop of honey or syrup is submerged in water or another miscible liquid, diffusion immediately smears out its interface, causing the drop to deform more easily in response to external forces (stirring or buoyant forces). As a result, surprising flow responses can emerge; For example, sessile drops have been found to form miscible "skirts" due to free convection[11], and pendant drops have been reported to produce a remarkable jet emanating from their apex[12].

Freely suspended, miscible drops can also display interesting behavior. Notably, Kojima et al.[13] studied the dynamics of syrup drops falling through syrup dilutions. They were especially interested in the shape transitions of these drops and present a theory to explain how an initially spherical drop develops into an open torus. Interestingly, and despite the fact that the fluids were fully miscible, they found that it was necessary to incorporate a small, but nonzero interfacial tension into their analysis to fully explain the shape changes seen in the experiments.

Vorobev et al.[14] consider the opposite case of rising miscible drops and use direct numerical simulations to assess the effect of the interfacial tension on the drop shape. They find that for very high interfacial tensions, the drop remains spherical, while for intermediate tensions, the rising drop deforms into a toroidal shape, recovering the behavior of falling drops reported by Kojima et al. Finally, for very small interfacial tensions, the drop is not able to maintain its shape and is dispersed into the surrounding liquid, with the notable exception of nearly density-matched mixtures, where the drop maintains a spherical shape without deforming.

Inspired by the above works, one of the authors of the present paper investigated buoyant water drops rising through syrup at low Reynolds number[15]. Using a syringe and needle to produce 1 -10 $\mu$L drops, and an optical setup to track their velocities and volumes, they found that these drops remain relatively spherical throughout their rise, suggesting that a finite interfacial tension stabilizes their shape. They also found the drops to display qualitatively different behavior depending on their travel time: On short time scales, the drops appeared to rise at constant velocities and volumes, while on long time scales, their volumes increased, and the velocities decreased. The authors suggested that the volumetric growth and the velocity decline were due to ambient liquid being swept into the drops from the back, but they did not attempt to verify this hypothesis using theoretical arguments; nor did they attempt to explain the observed power law behavior for velocity and volume or to predict the final drop size.

In order to address these limitations and to complement previous studies, we here present a unifying theory that predicts the velocities and volumes of miscible drops rising and falling through another fluid at low Reynolds number. Our predictions agree well with previously published experiments of water drops rising through syrup, and due to vanishingly small inertia and surface tensions, they are likely to find geophysical applications including buoyant plume dynamics in the Earth's mantle[16].

## 2. Model equations

In this work, we aim for a simple closed-form understanding of the drop dynamics. As such, we will leave the more general setting of continuum dynamics, and base our developments on the idealized geometric setting of a spherical drop. For the sake of nomenclature, we will refer to the two miscible fluids considered as "water" and "syrup", and use "rise" as the direction of motion of the drop.

Based on previous experimental observations[15], we make the following *a priori* modeling assumptions: (i) the drop can be well approximated as spherical; (ii) the time- and length-scales separate such that the internal liquid is well-mixed (spatially constant, but temporally variable, composition and density), buoyancy dominates diffusion at longer scales, thus as the drop rises, it is continuously exposed to syrup with constant (initial) composition and density, (iii) the transition region is of finite width, which is small relative to the size of the drop; (iv) convection is limiting for the mixing process, i.e. high Péclet number; (v) the density inside the drop is linearly proportional to the mass fraction of syrup; and (vi) the drop rises at low Reynolds numbers so that the viscous drag is proportional to velocity.

We emphasize that the combination of assumptions (ii) and (iii) leads to a model within the classical style of "sharp interface models", where the dynamics within the boundary layer is not explicitly represented in our model, but rather parameterized. This can be thought of as a multi-scale approach, where for the governing equations (mass calculations, dynamics, etc.), we consider the interface between syrup and water as being "sharp", with the boundary layer being so thin it does not impact the modeling. On the other hand, as we will see later, the finite width of the boundary layer enters the model through an effective diffusion rate.

We will divide our modeling in three main sections, reflecting the dominant processes in the system. In the "geometric relations" section, we will summarize the classical relations of density and volume as applied by the modeling assumptions (i), (ii), (iv) and (v). In the "hydrodynamics" section, we will summarize the relevant dynamics of low Reynolds number flows, consistent with modeling assumptions (i) and (vi). Finally, the most critical modeling choices are controlled by the mass exchange between the drop and the surrounding syrup, related to assumption (ii) and (iii), and this is developed in the "mass transfer" section.

### 2.1 Geometric relations

Subject to the modeling assumptions, we can describe the drop with its radius $R(t)$, as indicated in Figure 1. It follows from assumption (i) that the effect of the tail is negligible. We denote the density of pure water and syrup as $\rho_w^*$ and $\rho_s^*$, and the mass of water and syrup in the drop as $m_w^*(t)$ and $m_s^*(t)$, respectively. Here and in the following an asterisk denotes dimensional quantities which will later be non-dimensionalized, although to avoid unnecessary asterisks, we will not mark dimensional quantities (such as the radius $R(t)$) for which we do not require a non-dimensional counterpart.

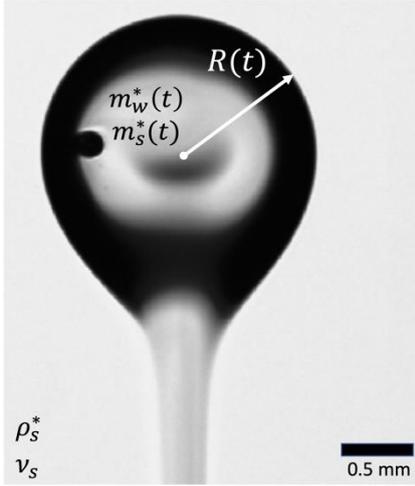

**Figure 1:** Properties of a freely suspended drop used in the mathematical model. $R(t)$ is the drop radius, $\rho_s^*$ is the density of pure syrup (ambient fluid), $\nu_s$ is the viscosity of pure syrup, and $m_w^*(t)$ and $m_s^*(t)$ denote the mass of water and syrup in the drop, respectively. The dark central spot is a reflection from the camera. The photo is taken from the same image data that was used to produce the experimental results published in Mossige et al., 2021[15].

The volume and surface area of a spherical drop are given by the standard expressions:

$$V^*(t) = \frac{4\pi}{3} R(t)^3; \qquad A(t) = 4\pi R(t)^2. \tag{2.1}$$

As stated in our modeling assumptions, and in particular as a consequence of (iii) and (iv), the volume can also be well approximated as a linear function of the mass of each component

$$V^*(t) = \frac{m_w^*(t)}{\rho_w^*} + \frac{m_s^*(t)}{\rho_s^*}. \tag{2.2}$$

We will also need the mixture density, which is given by the fraction of mass to volume,

$$\rho^*(t) = \frac{m_w^*(t) + m_s^*(t)}{V^*(t)}. \tag{2.3}$$

We use the initial drop mass $m_{w,0}^*$ and volume $V_0^* = m_{w,0}^*/\rho_w^*$ together with syrup density $\rho_s^*$ as characteristic values, to obtain the non-dimensionalized quantities

$$m_w(t) = \frac{m_w^*(t)}{m_{w,0}^*}, \qquad m_s(t) = \frac{\rho_w}{\rho_s} \frac{m_s^*(t)}{m_{w,0}^*}, \quad V(t) = \frac{V^*(t)}{V_0^*}, \text{ and } \rho_x = \frac{\rho_x^*}{\rho_s^*}. \tag{2.4}$$

with $x = [w, s, -]$, where "$-$" refers to the mixture density. The internal water concentration is given in terms of the density difference between the drop and its surroundings as:

$$x_w(t) = \frac{\rho_s^* - \rho^*(t)}{\rho_s^* - \rho_w^*} = \frac{1 - \rho(t)}{1 - \rho_w} \quad \text{with} \quad \rho(t) = \frac{\rho_w m_w(t) + m_s(t)}{V(t)}. \tag{2.5}$$

With this non-dimensionalization, the above dimensionless quantities satisfy the relations

$$V(t) = m_w(t) + m_s(t) \tag{2.6}$$

as well as

$$x_w(t) = \frac{1}{1-\rho_w} - \frac{\rho_w m_w(t) + m_s(t)}{(1-\rho_w)V(t)} \tag{2.7}$$

## 2.2 Hydrodynamics

We base our hydrodynamical considerations on low Reynolds number flow. In this setting, the viscous drag is dominated by the external fluid, and is linearly proportional to viscosity $v_s$ and velocity $U^*(t)$:

$$F_D(t) = c_d v_s \rho_s\, U^*(t) V^*(t)^{1/3}, \qquad (2.8)$$

which is simply the Stokes' drag law[17,18] with proportionality constant $c_d = 6\pi \left(\frac{3}{4\pi}\right)^{1/3}$. We disregard acceleration, as this is typically not relevant for low Reynolds numbers, and the viscous drag must then be balanced by the buoyant force associated with the lower density in the drop,

$$F_B(t) = \left(\rho_s^* - \rho^*(t)\right) g V^*(t), \qquad (2.9)$$

as shown in Figure 2a. Here $g$ is the gravitational constant. By equating the forces above and solving for the velocity one obtains

$$U^*(t) = \frac{(\rho_s^* - \rho^*(t)) g V^*(t)}{c_d v_s \rho_s^* V^*(t)^{1/3}}. \qquad (2.10)$$

We nondimensionalize this expression by introducing the characteristic velocity as the initial drop velocity as given by a water-filled drop of initial volume $V_0^*$:

$$U_0^* = \frac{(\rho_s^* - \rho_w^*) g (V_0^*)^{2/3}}{c_d v_s \rho_s^*}, \qquad U(t) = \frac{U^*(t)}{U_0^*}. \qquad (2.10)$$

In terms of dimensionless quantities, we obtain that the velocity is given as the product of concentration and the square of the cube root of volume,

$$U(t) = x_w(t) V(t)^{2/3}. \qquad (2.11)$$

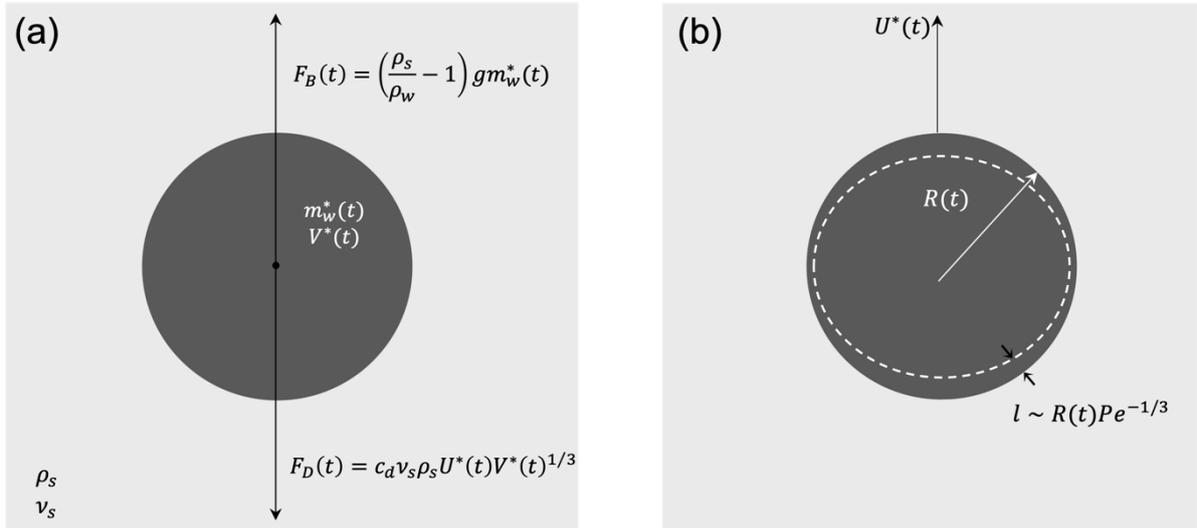

**Figure 2:** (a) Hydrodynamics: The buoyant force on a drop, $F_B$, is counteracted by the viscous drag force, $F_D$. The drag coefficient is given by $c_d = 6\pi(3/4\pi)^{1/3}$. (b) Mass transfer: When a drop with radius $R(t)$ translates through a viscous fluid at $Re \equiv U^*(t)R(t)/v_s^* \ll 1$ and $Pe \equiv U^*(t)R(t)/D^* \gg 1$, the thickness of the miscible layer enveloping the drop is given by $l \sim R(t)Pe^{-1/3}$, as indicated by the broken line. The thickness $l$ can depend on latitudinal position relative to the center of the drop, in response to the structure of the laminar flow field.

## 2.3 Mass transfer

In order to complete our mathematical model, we must also account for mutual diffusion between the drop and the surrounding fluid, which leads to a miscible boundary layer. If the drop were stationary, this layer would grow continuously in time, however, due to its upward motion, we expect a finite velocity-dependent boundary layer $\ell$, as illustrated in Figure 2b. Indeed, for small Reynolds numbers and large Péclet numbers, the thickness of the layer is given by[19]:

$$\ell(t) \sim R(t) Pe^{-1/3} = R(t)\left(\frac{D^*}{U^*(t)R(t)}\right)^{1/3} =$$
$$R(0)\left(\frac{R(t)}{R(0)}\right)^{2/3}\left(\frac{D^*}{U_0^* R(0)}\right)^{1/3} U(t)^{-\frac{1}{3}} \sim (V_0^*)^{1/3} D^{1/3} V(t)^{2/9} U(t)^{-\frac{1}{3}}, \qquad (2.12)$$

where the last $\sim$ is due to the proportionality between radius and cube root of volume. Furthermore, $D^*$ denotes the mutual diffusivity between water and syrup, which we non-dimensionalize in equation (2.12) as

$$D = \frac{D^*}{U_0^* R(0)}. \qquad (2.13)$$

Note that for this non-dimensionalization, the dimensionless mutual diffusivity corresponds to the inverse of the Péclet number at the initial time.

The thickness of the boundary layer can be used as a characteristic length scale for Fickian diffusion, such that for component species $\xi = \{w, s\}$ we obtain

$$j_\xi \sim D_\xi^* \frac{x_w(t)}{\ell(t)} \sim -(V_0^*)^{-1/3} D_\xi^* D^{-\frac{1}{3}} x_w(t) V(t)^{-\frac{2}{9}} U(t)^{\frac{1}{3}}. \tag{2.14}$$

Here $D_\xi^*$ is the effective diffusivity of component $\xi$ across the interface region. This parameter accounts for the diffusivity of water across the boundary layer being different to the diffusivity of syrup across the boundary layer due to the strong non-linear dependency of diffusivity on concentration in the glucose-water system[20]. It can in principle be calculated, and this calculation would be quite easy for a 1D problem. However, we are not aware of such effective values of diffusivity across the boundary layer having been reported for this more complex system. The sign convention is chosen to have positive diffusive fluxes out of the drop, so that $j_s$ is expected to be positive, while $j_w$ is expected to be negative. Diffusion has units of area per time, so that the diffusive flux $j_\xi$ has units of volume per area per time. We obtain mass fluxes for the drop by multiplying with the densities of the pure mixture and the area of the drop.

Summarizing these modeling considerations, the total diffusive mass exchange across the miscible boundary layer is approximated in terms of dimensional time $t^*$ as:

$$\frac{d}{dt^*} m_\xi \sim A(t^*) \rho_\xi j_\xi \sim \rho_\xi^* (V_0^*)^{1/3} D_\xi^* D^{-\frac{1}{3}} x_w(t^*) V(t^*)^{-\frac{2}{9}} U(t^*)^{\frac{1}{3}}. \tag{2.15}$$

Finally, we simplify this expression by introducing a proportionality constant, $\kappa$, which in addition to the geometric factors implied above reflects that the thickness of the boundary layer is variable on the drop surface:

$$\frac{d}{dt^*} m_w^*(t^*) = -\kappa \rho_w^* (V_0^*)^{1/3} D_w^* D^{-1/3} x_w(t^*) V(t^*)^{-2/9} U(t^*)^{\frac{1}{3}}, \tag{2.16a}$$

$$\frac{d}{dt^*} m_s^*(t^*) = \kappa \rho_s^* (V_0^*)^{1/3} D_s^* D^{-1/3} x_w(t^*) V(t^*)^{-2/9} U(t^*)^{\frac{1}{3}}. \tag{2.16b}$$

Equations (2.16) suggest the following characteristic and non-dimensional time $t$ as:

$$t_0^* = \frac{m_{w,0}^* D^{\frac{1}{3}}}{\kappa \rho_w^* (V_0^*)^{1/3} D_s^*} = \frac{(V_0^*)^{2/3} D^{\frac{1}{3}}}{\kappa D_s^*} \quad \text{and} \quad t = \frac{t^*}{t_0^*}. \tag{2.17}$$

For this choice equations (2.16) can be written in dimensionless form as:

$$\frac{d}{dt} m_w(t) = -x_w(t) V(t)^{4/9} U(t)^{\frac{1}{3}} \tag{2.18a}$$

$$\frac{d}{dt} m_s(t) = \mathcal{D} x_w(\tau) V(t)^{4/9} U(t)^{\frac{1}{3}}. \tag{2.18b}$$

We remark that the mass exchange ratio $\mathcal{D} = \frac{D_s^*}{D_w^*}$ is the amount of mass of syrup diffusing across the boundary layer per mass of water. As discussed after equation (2.14), this ratio is a reflection of the non-linear dependency of the diffusion coefficients on the concentration, and the actual concentration profile across the boundary layer. It will play an important role in the later development.

## 2.4 Summary of model equations.

We summarize the non-dimensional model equations as follows: The model is described by five time-dependent quantities, mass (within the drop) of water and syrup, $m_w(t)$ and $m_s(t)$, together with the volume of the drop $V(t)$, the concentration of water $x_w(t)$, and the velocity $U(t)$. These five quantities are subject to two volumetric constraints, given in equations (2.6) and (2.7), together with a force balance, given in equation (2.11), and two ordinary differential equations, given in equation (2.18).

As to the solvability of this system, we note that given the component masses $m_w(t)$ and $m_s(t)$ at any time $t$, the volumetric constraints and force balance immediately allow for the calculation of $V(t)$, $x_w(t)$ and $U(t)$. Furthermore, the dependence of these quantities on the component masses is smooth. These quantities can therefore formally be eliminated, reducing the system to two coupled non-linear ordinary differential equations for $m_w(t)$ and $m_s(t)$. The (local in time) solvability of this system follows from standard theory for ordinary differential equations.

# 3. Analysis of the drop rise dynamics

Section 2 outlines the governing equations for a drop rise within the physical regime under consideration. However, the presentation depends on several dimensionless quantities, and the resulting dynamics are not clear. In this section, we will show that the rise dynamics can be fully characterized by rather simple expressions.

In the first parts, section 3.1 and 3.2, we consider the general case of arbitrary $\mathcal{D}$, with the exception of the degenerate cases, i.e we assume $0 < [\mathcal{D} \neq 1] < \infty$. The particular degenerate cases $\mathcal{D} \in \{0, 1, \infty\}$ are discussed in section 3.3.

## 3.1 Species concentration, final size and velocity

As a preliminary calculation, we recognize that independent of the time-evolution of the above system, the final drop size can be directly characterized by pure mass balance arguments. This comes as a consequence of the mass exchange model, since by dividing equation (2.18) with itself (for the two choices $\xi = \{w, s\}$), we obtain that the two masses of the system are related by:

$$\frac{d}{dt}m_s = -\mathcal{D}\frac{d}{dt}m_w. \tag{3.1}$$

This equation can be integrated from $t = 0$, for which one obtains:

$$m_s(t) = -\mathcal{D}(m_w(t) - 1), \tag{3.2}$$

where we have used the initial conditions and non-dimensionalization which ensure that $m_s(0) = 0$ and $m_w(1) = 1$. In view of equation (2.6), this implies a linear relationship between mass of syrup and volume,

$$V(t) = m_w(t) + m_s(t) = (1 - \mathcal{D}^{-1})m_s(t) + 1. \tag{3.3}$$

Furthermore, at $t = \infty$ the drop will have equilibrated with the external syrup, and thus $m_w(\infty) = 0$. From equation (3.2) and (3.3), we therefore obtain the following relationships at final time:

$$m_{s,\infty} = \mathcal{D} = V_\infty. \tag{3.4}$$

With the above expressions, we can obtain after some algebraic manipulations an elegant expression for the concentration of water inside the drop, replacing the somewhat unsightly equation (2.7) (see appendix for derivation):

$$x_w(t) = \frac{1}{\mathcal{D}-1}\left(\frac{\mathcal{D}}{V(t)} - 1\right). \tag{3.5}$$

The above non-dimensionalizations and calculations allow us to also state the drop velocity only as a function of its volume, without explicit dependence on density. Indeed, equations (2.11) and (3.5) combine to yield:

$$U(t) = \frac{\mathcal{D}-V(t)}{\mathcal{D}-1} V(t)^{-1/3}. \tag{3.6}$$

### 3.2 General case of dynamical equations for drop size

From section 3.1, we recognize that a key parameter in the drop evolution is the mass exchange ratio $\mathcal{D}$, and that a convenient dimensionless quantity to characterize the system is the dimensionless volume $V(t)$. We therefore proceed to obtain an equation for the time-evolution of $V(t)$, eliminating the four other time-dependent variables mentioned in Section (2.4).

By Equation (3.3), and (2.18) we obtain:

$$\frac{d}{dt}V(t) = (1-\mathcal{D}^{-1})\frac{d}{dt}m_s(t) = (1-\mathcal{D}^{-1})x_w(t)V(t)^{4/9}U(t)^{\frac{1}{3}}. \tag{3.7}$$

We can now use equations (3.5) and (3.6) to eliminate $x_w(t)$ and $U(t)$, leading to

$$\frac{d}{dt}V(t) = \mathcal{D}^{-\frac{4}{3}}\left(\frac{\mathcal{D}-V(t)}{\mathcal{D}-1}\right)^{\frac{1}{3}}\left(1-\frac{V(t)}{\mathcal{D}}\right)\left(\frac{V(t)}{\mathcal{D}}\right)^{-\frac{2}{3}}. \tag{3.8}$$

We recognize that the fraction within the cube root is always positive, and thus:

$$\left(\frac{\mathcal{D}-V(t)}{\mathcal{D}-1}\right)^{\frac{1}{3}} = \mathcal{D}\left|1-\frac{V(t)}{\mathcal{D}}\right|^{1/3}|\mathcal{D}-1|^{-1/3}$$

This equation suggests introducing a new dimensionless time and volume, defined by:

$$\tau = \mathcal{D}^{-\frac{4}{3}}|\mathcal{D}-1|^{-1/3}t \quad \text{and} \quad W(t) = \frac{V(t)}{\mathcal{D}}. \tag{3.9}$$

With this variable choice, the dynamics of all drops covered by our modeling assumptions can be described by the single, parameter-free ordinary differential equation:

$$\frac{d}{d\tau}W(\tau) = (1-W(\tau))|1-W(t)|^{1/3}W(\tau)^{-\frac{2}{3}}, \tag{3.10}$$

subject to the initial condition:

$$W(0) = \mathcal{D}^{-1}. \tag{3.11}$$

We emphasize that although the process combines variable density, two diffusion rates and viscous flow, the dynamics of this full parameter space is solely defined by the single curve given implicitly by equation (3.10). Moreover, equation (3.10) has well-known implicit solutions in terms of the hypergeometric function $F_{2,1}$ for each of the two branches of the absolute value, which are given by:

$$W(\tau)^{\frac{5}{3}}F_{2,1}\left(\frac{4}{3},\frac{5}{3};\frac{8}{3};W(\tau)\right) = \mathcal{D}^{-\frac{5}{3}}F_{2,1}\left(\frac{4}{3},\frac{5}{3};\frac{8}{3};\mathcal{D}^{-1}\right) + \frac{5\tau}{3} \qquad \text{for } D > 1 \qquad (3.12a)$$

$$(W(\tau)-1)^{-\frac{1}{3}}F_{2,1}\left(-\frac{2}{3},-\frac{1}{3};\frac{2}{3};1-W(\tau)\right) = (\mathcal{D}^{-1}-1)^{-\frac{1}{3}}F_{2,1}\left(-\frac{2}{3},-\frac{1}{3};\frac{2}{3};1-\mathcal{D}^{-1}\right) + \frac{\tau}{3}$$
$$\text{for } D < 1. \qquad (3.12b)$$

This solution is universal for the problem, with the exception of three particular limit cases where the solution scaling fails to be valid. These can be seen most clearly from equation (3.9), which is invalid when $\mathcal{D}$ takes the values of $\{0, 1, \infty\}$. We consider these special cases in the next section.

### 3.3 Special cases in the limits of one-sided and balanced diffusion

For three special cases, the dimensionless time given in equation (3.9) fails to be meaningful. Firstly, these are the two limits where either $\mathcal{D} \to \infty$ or $\mathcal{D} \to 0$. Since we have established that $V_\infty = \mathcal{D}$, in these limits the drop either expands indefinitely, or dissolves completely, respectively. We also need to consider the case of balanced diffusion, i.e. $\mathcal{D} = 1$, for which the drop size is constant.

Considering first the case where diffusion into the drop dominates, $\mathcal{D} \to \infty$, we proceed from equation (3.8) to deduce the limit equation for $\overset{\infty}{V}$ (we denote the special cases by the value of $\mathcal{D}$ above the variable):

$$\frac{d}{dt}\overset{\infty}{V}(t) = \lim_{\mathcal{D}\to\infty} \mathcal{D}^{-\frac{1}{3}}(\mathcal{D}-1)^{-\frac{1}{3}}\left(1-\frac{V(t)}{\mathcal{D}}\right)^{\frac{4}{3}}\left(\frac{V(t)}{\mathcal{D}}\right)^{-\frac{2}{3}} = \overset{\infty}{V}(t)^{-\frac{2}{3}}. \qquad (3.13)$$

This equation can be integrated directly to obtain (including initial condition):

$$\overset{\infty}{V}(t) = \left(1+\frac{5}{3}t\right)^{\frac{3}{5}}. \qquad (3.14)$$

Equation (3.14) can also be obtained directly as the limit of (3.11) as $\mathcal{D} \to \infty$, since[21]

$$\lim_{\mathcal{D}\to\infty} F_{2,1}\left(\frac{4}{3},\frac{5}{3};\frac{8}{3};\mathcal{D}^{-1}\right) = 1 \qquad (3.15)$$

Considering now the case where diffusion out of the drop dominates, $\mathcal{D} \to 0$, equation (2.17) indicates that with the choice of dimensionless time used above, the drop instantly disappears. As such, we will for this special cause use a different dimensionless time, defined by $\tilde{t} = t/\mathcal{D}$, and proceed from equation (3.7) to deduce the limit equation for $\overset{0}{V}$:

$$\frac{d}{d\tilde{t}}\overset{0}{V}(\tilde{t}) = \frac{d}{d\mathcal{D}^\epsilon t}\overset{0}{V}(\tilde{t}) = \lim_{\mathcal{D}\to 0} \mathcal{D}\mathcal{D}^{\frac{2}{3}}(\mathcal{D}-1)^{-\frac{1}{3}}\left(1-\frac{V(\tilde{t})}{\mathcal{D}}\right)^{\frac{4}{3}}\left(\frac{V(\tilde{t})}{\mathcal{D}}\right)^{-\frac{2}{3}}$$

$$= \lim_{\mathcal{D}\to 0} \mathcal{D}^{\frac{2}{3}}\mathcal{D}^{-\frac{4}{3}}(\mathcal{D}-V(\tilde{t}))^{\frac{4}{3}}V(\tilde{t})^{-\frac{2}{3}}\mathcal{D}^{2/3} = -\overset{0}{V}(\tilde{t})^{\frac{2}{3}}. \qquad (3.16)$$

This equation can also be integrated exactly to yield:

$$\overset{0}{V}(\tilde{t}) = \left(1 - \tfrac{1}{3}\tilde{t}\right)^3. \tag{3.17}$$

The final limit case to consider is that of balanced diffusion, where $\mathcal{D} = 1$. By definition, the volume of the drop will now remain constant, $\overset{1}{V}(t) = 1$, which is also consistent with equation (3.7). It remains to consider the evolution of the drop density, since equation (3.5) fails to be valid. Proceeding therefore from equations (2.18b) and (2.11) we now obtain:

$$\tfrac{d}{dt}\overset{1}{m}_s(t) = \overset{1}{x}_w(t)^{4/3}. \tag{3.18}$$

Further, equations (2.6) and (2.7) imply

$$\tfrac{d}{dt}\overset{1}{x}_w(t) = -\tfrac{d}{dt}\overset{1}{m}_s(t), \tag{3.18b}$$

to yield for the concentration of water in this limit case:

$$\tfrac{d}{dt}\overset{1}{x}_w = -\overset{1}{x}_w(t)^{4/3}. \tag{3.19}$$

Which as above can be integrated exactly to yield:

$$\overset{1}{x}_w = \left(1 + \tfrac{1}{3}t\right)^{-3}. \tag{3.20}$$

### 3.4 Summary of the rise dynamics

We summarize the findings as follows. The general case considers the situation where the relative water loss from the drop compared to the gain of syrup, $\mathcal{D}$, is finite. The resulting system can be expressed in terms of a one-parameter family of solutions, defined in terms of $\mathcal{D}$, based only on the finite drop size:

1. The maximum drop size is defined solely on the mass exchange coefficients as given in equation (3.4).
2. The velocity and volume are related by equation (3.6).
3. The dimensionless volume satisfies an expression based on the hypergeometric function, as given in equation (3.12).

All observations for the general case can be extended to the three special cases of $\mathcal{D} \in \{0, 1, \infty\}$. Indeed, the special cases are simpler than the general case, and in particular the dimensionless volume can be given by closed-form expressions such as equation (3.14), (3.17) and (3.20). These closed-form expressions indicate the following scaling laws:

4. For $\mathcal{D} = \infty$, the late-time drop size satisfies $\overset{\infty}{V}(t) \sim t^{3/5}$ with velocity $\overset{\infty}{U}(t) \sim t^{-1/5}$
5. For $\mathcal{D} = 1$, the drop size is constant $\overset{1}{V} = 1$, with late-time drop concentration $\overset{1}{x}_w(t) \sim t^{-3}$ and velocity $\overset{1}{U}(t) \sim t^{-3}$.
6. For $\mathcal{D} = 0$, the drop size $\overset{0}{V}$ reaches 0 in finite time.

These dynamics are illustrated in Figure 3. Here, we show the evolution of drop size for eight different values of $\mathcal{D}$. The limit cases of $\mathcal{D} = \{1, \infty\}$ are shown in solid lines, while the intermediate cases $\mathcal{D} = \{10^{-2}, 10^{-1}, 0.5, 2, 10^{-1}, 10^{-2}\}$ are shown in dotted lines. On the chosen time-scales, the case $\mathcal{D} = 0$ implies instantaneous dissolution, and is therefore not shown. The left figure illustrates the results in the "natural" dimensionless timescale $t$ given by equation (2.17) and volume $V$. The right figure shows the same data with the time-scale $\tau$ defined in equation (3.9) and the rescaled dimensionless volume $W$. Note in particular that in this latter case, all the curves converge to $W = 1$.

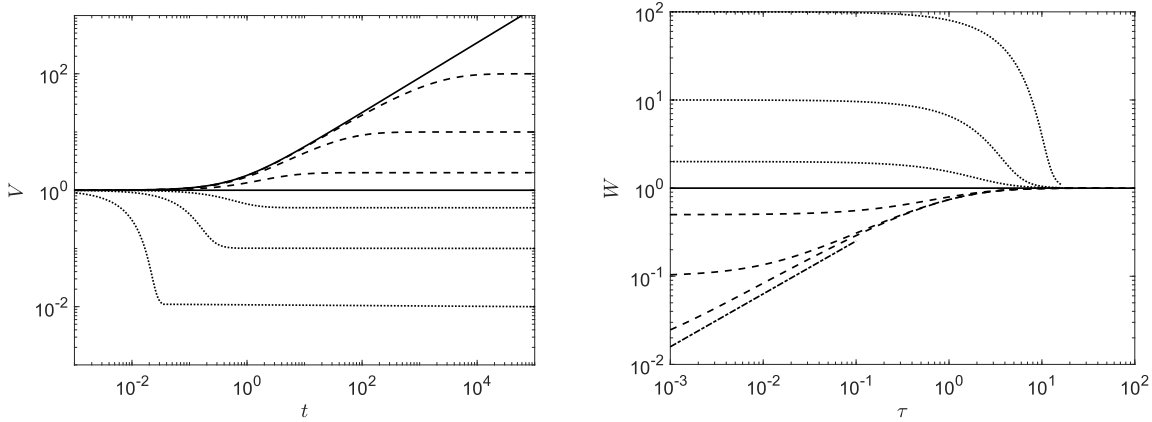

**Figure 3**: Left: Drop volume $V$ as a function of dimensionless time $t$ for values $\mathcal{D} = \{10^{-2}, 10^{-1}, 0.5, 0, 2, 10^1, 10^2, \infty\}$ (curves shift up with increasing $\mathcal{D}$, dots indicate $\mathcal{D} < 1$ while dashes indicate $\mathcal{D} > 1$). Right: same data as in left figure, plotted in terms of $W$ as a function of dimensionless time $\tau$ (curves shift down with increasing $\mathcal{D}$). These dimensionless quantities are not strictly defined for $\mathcal{D} = \infty$, however the $W \sim \tau^{3/5}$ scaling implied by equation (3.14) is nevertheless indicated by a dash-dotted line.

Of interest is also the drop velocity $U(t)$, as well as the vertical position of the drop $z(t) = \int_0^t U(t')dt'$. For $\mathcal{D} \neq 1$, the velocity $U(t)$ is plotted in Figure 4 (left) according to the expression given in equation (3.6) for the volumes $V(t)$ plotted in Figure 3 (left). The special case $\mathcal{D} = 1$ is also plotted by substituting equation (3.20) in equation (2.11). The position $z(t)$, is shown in Figure 4 (right) based on standard numerical integration of $U(t)$.

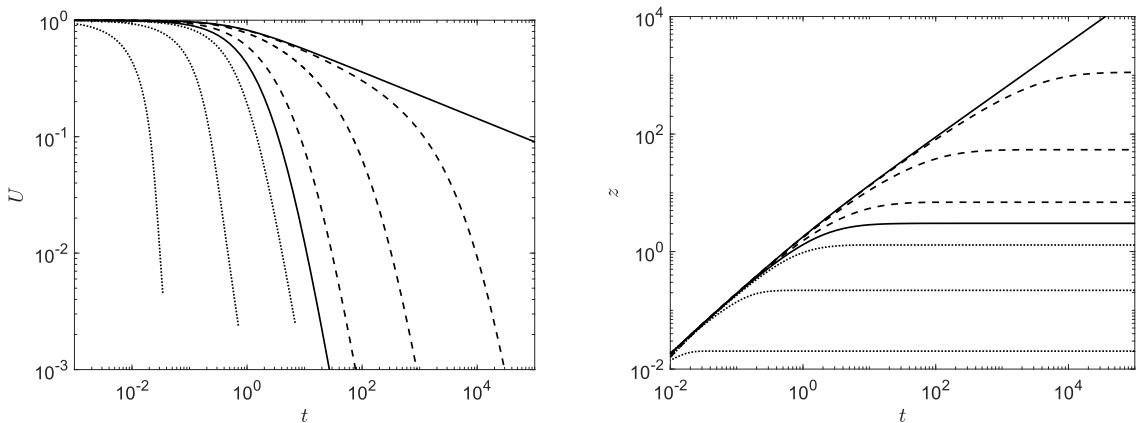

**Figure 4**: Left: Drop velocity $U$ as a function of dimensionless time $t$ for values $\mathcal{D} = \{10^{-2}, 10^{-1}, 0.5, 0, 2, 10^1, 10^2, \infty\}$ (curves shift right with increasing $\mathcal{D}$, dots indicate $\mathcal{D} < 1$ while dashes indicate $\mathcal{D} > 1$). Right: Vertical drop position $z = \int_0^t U(t')dt'$ as a function of dimensionless time $t$ (curves shift up with increasing $\mathcal{D}$).

# 4. Comparison with experimental data

In order to validate the mathematical model, we used published data for the volumes and velocities of water drops in syrup (see Ref. 15 and 22). The experiments were based on de-ionized water (1-10 $\mu$L; Milli-Q Academic A10) rising in corn syrup (Karo; light corn syrup) using an optical setup consisting of a collimated light source (telecentric lens: Model: 63074, Edmund Optics, NJ, USA; fiber optic light: Model: 21AC fiber optic illuminator, Edmund Optics, NJ, USA; collimator: Model: 62760, Edmund Optics, NJ, USA) and a back lit camera (Model: GPF 125C IRF, Allied Vision Technologies, PA, USA). The camera and light source were mounted onto a programmable stage (Model: ULM- TILT, Newport, CA, USA), which enabled the drop to be followed throughout its rise and its volume and velocity to be extracted.

In a typical experiment, syrup was first poured into a homemade glass chamber (50mm by 50mm wide, 150mm tall) and placed in a vacuum chamber to remove any air bubbles resulting from the filling procedure. Then, a water drop was injected into the bottom of the chamber though a self-healing membrane using a syringe (10 $\mu$l, Hamilton Microliter™ #801) and stainless-steel needle (OD = 0.362 mm). Due to vanishingly small capillary pressures between the miscible liquids, the buoyant drop starts to rise immediately after the injection. During the entire experiment, which typically took one hour, the drop is reported to remain spherical, with a distinct liquid-liquid interface. The experimental results and procedure are presented previously in Mossige et al[15].

Table 1 gives an overview of the physical properties of the fluids used in the experiments.

| Fluid property | Diffusivity water/syrup | Water density | Syrup density | Syrup viscosity |
|---|---|---|---|---|
| Symbol (units) | $D^*$ ($m^2/s$) | $\rho_w^*$ ($kg/m^3$) | $\rho_s^*$ ($kg/m^3$) | $\nu_s$ ($m^2/s$) |
| Value | $1.3 \cdot 10^{-10}$ [23] | 997 | 1,386 | $3.7 \cdot 10^{-3}$ |

**Table 1:** Physical properties of the experimental fluids. The viscosity of syrup was obtained using standard cone plate rheometry, as described in Ref. 15.

Our theoretical results indicate that the evolution of the drop depends on a single free parameter $\mathcal{D}$, representing the relative (effective) ratio of mass transfer between the syrup and the drop. As this parameter was not inferred in previous experiments, we show in the following a range of values $\mathcal{D} = \{5, 10, 20, \infty\}$. Additionally, we will use the experiments to identify the pair of proportionality constant $\kappa D_s^*$ entering the definition of dimensionless time, which represents the interplay between the flow field around the drop and the effective diffusion across the flow.

The evolution of volume over time for finite $\mathcal{D}$ is given by equation (3.12) and is plotted in Figure 5 (dotted lines), together with the same experimental data. The approximate solution given by Eq. 3.14 is shown by the solid line. The experimental data represent several different experiments with different initial volumes, and collapse given the non-dimensionalizations provided.

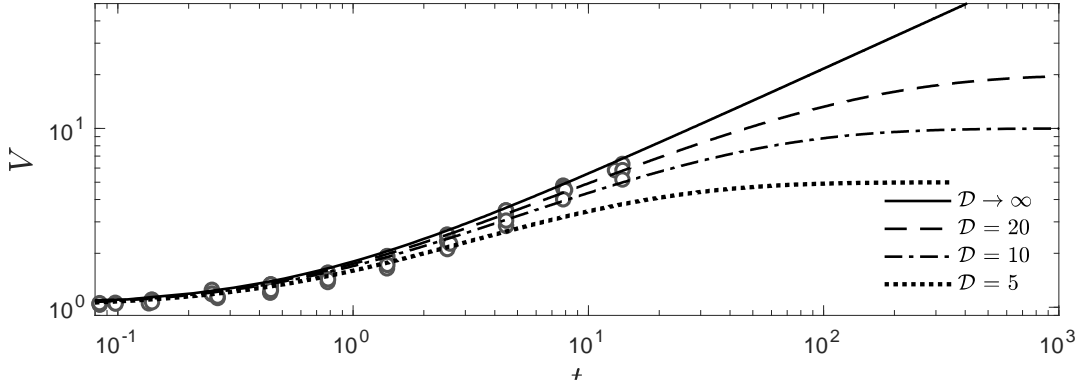

**Figure 5.** Dimensionless volume, $V$, plotted versus dimensionless time, $t$, for $\mathcal{D} = \{5, 10, 20, \infty\}$, where $\mathcal{D} = \{5, 10, 20\}$ corresponds to Eq. 3.11 and the approximation $\mathcal{D} \to \infty$ corresponds to Eq. 3.14. The data points represent volumetric data for water drops rising through syrup from Ref. 15.

The relationship between drop velocity and volume is given by equation (3.6). In Figure 6, we show the predicted velocity-volume curves together with experimental data (symbols).

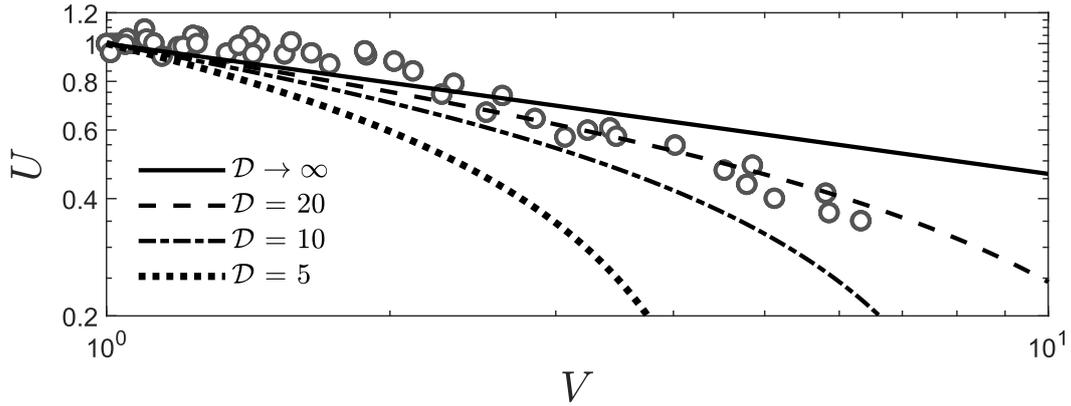

**Figure 6:** Dimensionless velocity, $U$, plotted versus dimensionless volume, $V$, for $\mathcal{D} = \{5, 10, 20, \infty\}$ (eqn. 3.6). The symbols represent experimental data for water drops rising through syrup from Ref. 15.

The original data was non-dimensionalized using the characteristic time

$$t^\dagger = \frac{(\nu_s \rho_s^*)^{2/3}}{(\rho_s^* - \rho_w^*)^{2/3} g^{2/3} (D^*)^{1/3}}. \tag{4.1}$$

In plotting figures 5 and 6, we heuristically identified the relationship between this dimensionless time and the dimensionless time stated in equation (2.17):

$$t^\dagger = 3 t_0^*. \tag{4.2}$$

Based on this observation, we propose the following expression for the effective mass diffusivity across the interface region $\kappa D_s^*$:

$$\kappa D_s^* = 3\left(\frac{3c_d^3}{4\pi}\right)^{\frac{1}{9}} \left(\frac{D^*(V_0^*)^{2/3}}{t^\dagger}\right)^{\frac{1}{2}} \tag{4.3}$$

This expression suggests the dependency between the drop scale parameters $\kappa$ and $D_s^*$, on the actual fluid dynamics and local-scale diffusion mechanisms (which exist in a more fine-grained modeling framework where concentration is represented as a continuous field variable). In particular, we propose

$$\kappa = 3\left(\frac{3c_d^3}{4\pi}\right)^{\frac{1}{9}} \quad \text{and} \quad D_s^* = \left(\frac{D^*(V_0^*)^{2/3}}{t^\dagger}\right)^{\frac{1}{2}} \tag{4.4}$$

The validity of these postulated relationships is supported by the agreement with experimental data shown in figures 5 and 6, however, it can be further strengthened in future work by comparison to direct numerical simulation of the coupled fluid-dynamics and mass conservation (diffusion) equations.

Based on the provided experimental data and our analysis, we infer that a mass transfer ratio of $\mathcal{D} \in \{10, 20\}$ seems to provide a reasonable match between theory and experiment. The main deviation is seen at early time in Figure 6 (close to $V = 1 = U$), and may be attributable to early-time effects in the experiment before the steady flows assumed in Section 2.2 have developed (note that dimensionless velocities higher than 1 are reported, which clearly indicates a non-steady regime).

Considering the above identified value of $\mathcal{D}$, our theory thus leads to a prediction of a final drop size on the order of 10 to 20 times the initial size. This prediction could be verified given a sufficiently tall experimental chamber, which according to Figure 4 (right) would require an experimental setup of a dimensionless height $z_\infty = z(t = \infty)$ of about 50 to 100. The non-dimensionalization of height follows from the definition of dimensionless time and velocity, thus using the upper estimate of $z_\infty \approx 100$ and $\mathcal{D} \approx 20$, in terms of physical quantities we predict the need for an experimental column of height

$$z_\infty^* = z_\infty \, U_0^* t_0^* = z_\infty \frac{\mathcal{D}^{1/3}}{3\left(\frac{3}{4\pi}\right)^{\frac{1}{9}}} \left(\frac{(\rho_s^* - \rho_w^*)g}{c_d^2 v_s \rho_s^* D^*}\right)^{2/3} V_0^* \,. \tag{4.5}$$

For the properties and drop size used in this experiment, this implies a column height on the order of 30 meters should be sufficient to capture the whole rise of the drop. As indicated from expression (4.5), more practical column heights can be achieved by modifying the fluid properties, or by initializing the experiment with drops of smaller volumes.

# 5. Conclusion

In this work, we explore the dynamics of a buoyant drop rising or falling through a miscible fluid at low Reynolds number. We present a theoretical model for velocity and volume, wherein the general case diffusion allows mass to both enter and leave the drop. Additionally, three special cases are treated. The first set considers negligible mass loss from the drop, which causes it to expand indefinitely, and the second set describes the opposite case where diffusion out of the drop dominates, which causes it to dissolve completely. The third and final case is that of balanced diffusion, where the drop volume remains constant.

Our theoretical calculations agree well with experimental data for millimeter sized water drops rising through syrup throughout the whole time-series of the experiment, corresponding to mass exchange ratios between 10 and 20, contrasting previous analyses which focused on identifying empirical scaling laws. Our model further directly leads to predictions of the final drop size and vertical position, not available in current experimental literature. Furthermore, the comparison to experimental data allows us to postulate the effective mass transfer parameters which appear in our theory.

# Acknowledgements

The authors would like to thank Bo Guo for helpful comments on the manuscript. EJLM would like to thank Prof. Gerald G. Fuller, Dr. Daniel J. Walls and Dr. Vineeth 'Vinny' Chandran Suja for a productive experimental investigation on rising miscible drops during his postdoctoral studies at Stanford University, which laid the basis of the current theoretical study.

# Appendix

This appendix gives the derivation of equation (3.5). Based on the simplicity of the initial and final expressions, the authors suspect that a more concise and elegant derivation exists, however, in lieu of this, the below is provided. Starting from equation (2.7),

$$x_w(t) = \frac{1}{1-\rho_w} - \frac{\rho_w m_w(t) + m_s(t)}{(1-\rho_w)V(t)}$$

Eliminating first $m_w(t)$ using equation (3.2):

$$x_w(t) = \frac{1}{1-\rho_w} - \frac{\rho_w\big(1-\mathcal{D}^{-1}m_s(t)\big) + m_s(t)}{(1-\rho_w)V(t)} = \frac{1}{(1-\rho_w)V(t)}\big(V(t) - \rho_w - (1-\mathcal{D}^{-1}\rho_w)m_s(t)\big)$$

Now further eliminating $m_s(t)$ using equation (3.3):

$$x_w(t) = \frac{1}{(1-\rho_w)V(t)}\left(V(t) - \rho_w - (1-\mathcal{D}^{-1}\rho_w)\frac{V(t)-1}{1-\mathcal{D}^{-1}}\right)$$

Introducing a common factor and collecting terms:

$$x_w(t) = \frac{V(t)(1-\mathcal{D}^{-1}) - \rho_w(1-\mathcal{D}^{-1}) - (1-\mathcal{D}^{-1}\rho_w)(V(t)-1)}{(1-\rho_w)V(t)(1-\mathcal{D}^{-1})}$$

$$= \frac{V(t) - V(t)\mathcal{D}^{-1} - \rho_w + \mathcal{D}^{-1}\rho_w - V(t) - \mathcal{D}^{-1}\rho_w + \mathcal{D}^{-1}\rho_w V(t) + 1}{(1-\rho_w)V(t)(1-\mathcal{D}^{-1})}$$

$$= \frac{-V(t)\mathcal{D}^{-1} - \rho_w + \mathcal{D}^{-1}\rho_w V(t) + 1}{(1-\rho_w)V(t)(1-\mathcal{D}^{-1})} = \frac{-V(t)\mathcal{D}^{-1} + 1}{V(t)(1-\mathcal{D}^{-1})} = \frac{-1 + V(t)^{-1}}{\mathcal{D} - 1}$$

The final expression corresponds to equation (3.5).